\documentclass[aps,twocolumn,showpacs,floats,prl]{revtex4}
\usepackage{graphicx}
\usepackage{dcolumn}
\usepackage{bm}
\usepackage{color}
\usepackage{subfigure}

\begin{document}

\author{N. Matveeva}
\author{S. Giorgini}
\affiliation{Dipartimento di Fisica, Universit\`a di Trento and CNR-INO BEC Center, I-38050 Povo, Trento, Italy}

\title{The impurity problem in a bilayer system of dipoles} 

\begin{abstract} 
We consider a bilayer geometry where a single impurity moves in a two-dimensional plane and is coupled, via dipolar interactions, to a two-dimensional system of fermions residing in the second layer. Dipoles in both layers point in the same direction oriented by an external field perpendicular to the plane of motion. We use quantum Monte Carlo methods to calculate the binding energy and the effective mass of the impurity at zero temperature as a function of the distance between layers as well as of the in-plane interaction strength. In the regime where the fermionic dipoles form a Wigner crystal, the physics of the impurity can be described in terms of a polaron coupled to the bath of lattice phonons. By reducing the distance between layers this polaron exhibits a crossover from a free-moving to a tightly-bound regime where its effective mass is orders of magnitude larger than the bare mass.     
\end{abstract}

\maketitle 

The polaron problem, in the broad sense of an impurity coupled to a bath of elementary excitations, is of general and fundamental interest in condensed matter physics. The original formulation of the polaron model addressed the motion of electrons coupled to
 the lattice vibrations of a crystal~\cite{MahanBook}. In this context phonon excitations are found to dress the impurity, thereby increasing its effective mass. For strong interactions, self-trapping of polarons was predicted as a result of the dragged phonon 
cloud which creates a confining potential where the impurity is finally trapped~\cite{Landau46}. Since the variational calculation by Landau and Pekar, the polaron problem in the strong-coupling regime has been investigated using many different theoretical tools
~\cite{Devreese09}, including exact quantum Monte Carlo (QMC) methods~\cite{Alexandrou90, Prokofev98, Wang98, Titantah01,Fantoni12}. On the experimental side, clear evidence of the self-trapping of polarons is still lacking, also due to the difficulty of accessing large interaction strengths in solid-state devices~\cite{MahanBook}.

An important recent extension of the polaron concept concerns the field of ultracold atoms. A polaron-like behavior is indeed expected from an impurity immersed in a Bose-Einstein condensate, which provides the phonon modes of the bath~\cite{Cucchietti06, Kalas06,
 Tempere09}, as well as in a Fermi sea, where the excitations of the medium have fermionic nature. This latter case is particularly interesting since both attractive~\cite{Lobo06, Chevy06, Prokofev08} and repulsive~\cite{Pilati10, Massignan11, Schmidt11} polarons
 have been considered theoretically and characterized in experiments~\cite{MIT09, Grimm12, Kohl12}. The quantum dynamics of an impurity in a Bose gas has also been recently observed~\cite{Catani12, Bloch13}.

In this Letter we propose a realization of the polaron model by using an impurity coupled via dipolar interactions to a two-dimensional (2D) system of fermions in a bilayer geometry~\cite{Note}. By tuning the in-plane dipolar interaction strength the state of 
the fermions can be turned from the Fermi liquid (FL) to the Wigner crystal (WC) phase, thereby changing the nature of the elementary excitations of the bath from fermionic to bosonic. QMC simulations are performed to calculate the binding energy and the effective 
mass of the impurity as a function of the distance between layers assuming that the interlayer potential barrier is high enough to suppress tunneling. In the WC phase the impurity exhibits a crossover from a free-moving to a tightly-bound regime similar to the self-trapping transition. However, in contrast to the paradigmatic case of electrons in a crystal, the coupling to phonons is found to decrease the effective mass of the impurity with respect to the band mass determined by the static periodic potential and to favor hopping processes of the impurity between lattice sites.  

We consider a system of $N+1$ identical dipolar fermions of mass $m$ and dipole moment $d$. $N$ fermions occupy the first layer (bottom layer) and the extra fermion occupies alone the second layer (top layer). The layers are 2D parallel planes separated by a distance $\lambda$ and motion in the transverse direction is completely frozen by a strong confining potential provided, for example, by an intense optical lattice. Since interlayer tunneling is assumed to be completely suppressed, the particle in the top layer can be considered as a distinguishable impurity coupled via dipolar interactions to the particles of the bottom layer. An external field aligns the dipoles in the direction perpendicular to the layers, so that in-plane interactions are purely repulsive and scale with the interparticle distance as $1/r^3$ while the interlayer particle-impurity potential is given by
\begin{equation}
V(r_{ai}) = \frac{d^2(r_{ai}^2 - 2 \lambda^2)}{(r^2_{ai}+\lambda^2)^{5/2}} \;,
\label{V} 
\end{equation}
where $r_{ai}=|{\bf r}_i-{\bf r}_a|$  is the in-plane distance between the $i$-th particle and the projection of the impurity position onto the bottom layer. The full Hamiltonian of the system is written as 
\begin{equation}
H=-\frac{\hbar^2}{2m}\sum_{i=1}^N\nabla_i^2+\sum_{i<j}\frac{d^2}{r_{ij}^3} -\frac{\hbar^2}{2m}\nabla_a^2+\sum_{i=1}^NV(r_{ai}) \;.
\label{Hamilton}
\end{equation}
Here $r_{ij}$ is the distance between a pair of particles in the bottom layer and $-(\hbar^2/2m)\nabla_a^2$ is the kinetic energy of the impurity in terms of its projected coordinate ${\bf r}_a$. 
The strength of the in-plane and interlayer dipolar interaction is expressed in terms of the dimensionless parameters $k_Fr_0$ and $k_F\lambda$, respectively. The characteristic length $r_0=md^2/\hbar^2$ arises from the dipole-dipole force and $k_F=\sqrt{4\pi n}$ is the Fermi wavevector of a 2D gas determined by the density $n$ in the bottom layer. The corresponding Fermi energy is given by $\epsilon_F=\hbar^2k_F^2/(2m)$. A similar impurity problem in a bilayer configuration was considered in Ref.~\cite{Klawunn13} in the limit of weak interactions in the bottom layer.

\begin{figure}
\begin{center}
\includegraphics[width=8.5cm]{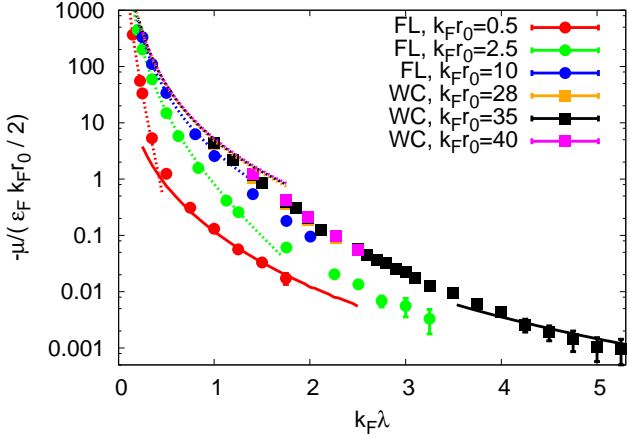}
\caption{(color online). Binding energy of the impurity as a function of the distance between layers for different values of the in-plane interaction strength $k_Fr_0$. Circles refer to the FL and squares to the WC phase. Dashed lines correspond to the binding energy of the two-body problem calculated from the Schr\"odinger equation. Solid lines are the results of perturbation theory holding when $k_F\lambda$ is large. The red line refers to a weakly interacting FL state, while the black line corresponds to Eq.~(\ref{perturbation1}) in the WC phase.}
\label{fig1}
\end{center}
\end{figure}

The equation of state of the single-layer 2D dipolar Fermi gas at $T=0$ was calculated using the fixed-node diffusion Monte Carlo (FNDMC) method~\cite{Matveeva12}. A phase transition from a Fermi liquid (FL)  to a Wigner crystal (WC) is predicted at the critical interaction strength $(k_Fr_0)_c=25\pm3$. Here, we use the same numerical technique to calculate the binding energy of the impurity, defined as the energy difference between the ground state with and without the impurity, $\mu=E_{N+1}-E_N$, and its effective mass. The latter is obtained from the diffusion coefficient of the impurity in imaginary time $\tau$~\cite{Boninsegni95,Boronat99}  
\begin{equation}
\frac{m}{m^*}=\lim_{\tau \rightarrow \infty} \frac{\langle| \Delta {\bf r}_a(\tau) |^2\rangle}{4D\tau} \;, \label{Diff} 
\end{equation}
where $D=\hbar^2/2m$ is the diffusion constant of a free particle and  $\langle| \Delta {\bf r}_a(\tau)|^2\rangle=\langle|{\bf r}_a(\tau)-{\bf r}_a(0)|^2\rangle$ is the mean square displacement of the impurity. Simulations are carried out both in the WC phase [$k_Fr_0>(k_Fr_0)_c$] and in the FL phase [$k_Fr_0<(k_Fr_0)_c$].

The technical details of the simulations are similar to the ones reported in Ref.~\cite{Matveeva12}. To simulate the bottom layer we use a box of volume  $\Omega=L_xL_y$, with $L_x\leq L_y$, and periodic boundary conditions (PBC) in both spatial directions.
The fermionic density in the bottom layer is $n=N/\Omega$. The calculation of the in-plane and interlayer dipolar interaction energy are performed by considering replicas of the simulation box and by carrying out the summation over pairs of particles with separation up to the cut-off distances
 $R_{c_1}=0.5L_x$ and $R_{c_2}=2L_x$, respectively~\cite{Supplemental}. The contribution from distances larger than $R_{c_1}$ ($R_{c_2}$) is accounted for by assuming a uniform distribution of particles which yields the tail energy 
$E_{\text{tail}_1}=\pi n d^2/R_{c_1}$ ($E_{\text{tail}_2} = 2\pi n R_{c_2}^2/(\lambda^2+R_{c_2}^2)^{3/2}$). We checked that larger values of $R_{c_1}$ and $R_{c_2}$ give the same $\mu$ and $m^*$ within statistical uncertainty. We notice that $R_{c_1}$ is
 significantly smaller than $R_{c_2}$ because in-plane interactions largely cancel when the difference $E_{N+1} - E_N$ is considered. Calculations were performed with different numbers of particles, ranging from $N=30$ to $N=90$ in the WC and from $N=29$ to $N=61$ 
in the FL phase, and no appreciable finite-size dependence is found for the binding energy and the effective mass. All results reported in this Letter are obtained using $N=30$ in the WC phase and $N=29$ in the FL phase.

\begin{figure}
\begin{center}
\includegraphics[width=8.5cm]{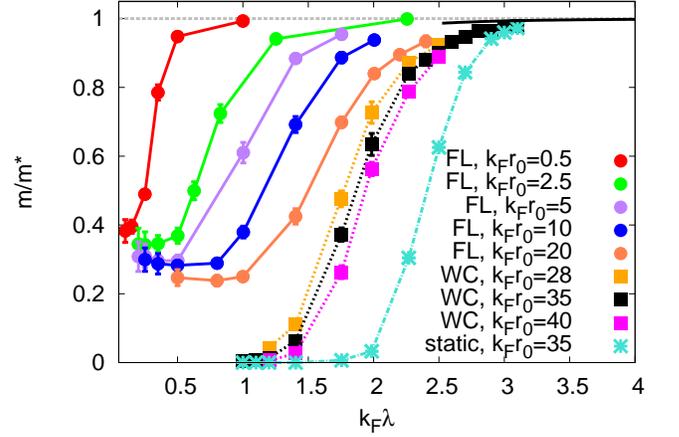}
\caption{(color online). Inverse effective mass of the impurity as a function of the distance between layers for different values of the in-plane interaction strength $k_Fr_0$. Circles refer to the FL and squares to the WC phase. The stars correspond to $m/m^\ast$ when only the static periodic potential $U({\bf r}_a)$ is considered and phonons are frozen. Lines connecting the symbols are a guide to the eye. The solid black line corresponds to Eq.~(\ref{perturbation2}) in the WC phase.}
\label{fig2}
\end{center}
\end{figure}

The FNDMC method is based on the choice of a trial wavefunction giving the many-body nodal surface which is kept fixed during the simulation~\cite{Kolorenc11}. We use a Jastrow-Slater function of the form
\begin{equation}
 \psi_T({\bf r}_1,...,{\bf r}_N,{\bf r}_a)=\prod_{i=1}^N h(r_{ai}) \prod_{i<j}f(r_{ij})\det[\varphi({\bf r}_i)]\;,
\label{Trial}
\end{equation}
where $h(r)$ and $f(r)$ are two-body non-negative correlation terms and the nodes are determined by the antisymmetric Slater determinant of single-particle orbitals $\varphi({\bf r})$. In the FL phase we use plane waves:
 $\varphi({\bf r})=e^{i{\bf k}_\alpha\cdot{\bf r}}$,  where ${\bf k}_\alpha=(2\pi/L)(n_\alpha^x,n_\alpha^y)$ are the wavevectors complying with PBC in the square box of size $L$. In the WC phase, instead, we use Gaussians: $\varphi({\bf r})=e^{-({\bf r}-{\bf R}_m)^2/\alpha^2}$ tied to the points ${\bf R}_m=(m_x+\frac{1}{2}m_y)a\hat{{\bf x}}+m_y\frac{\sqrt{3}}{2}a\hat{{\bf y}}$ of the triangular Bravais lattice in the $x-y$ plane. Here, $m_{x,y}$ are integers, $a=\sqrt{8\pi/\sqrt{3}}/k_F$ is the lattice spacing and $\alpha$ is a variational parameter to be optimized~\cite{Matveeva12}.
The Jastrow terms $h(r)$ and $f(r)$ are introduced to reduce the statistical variance and describe the correlations arising, respectively, from interlayer and in-plane dipolar interactions. The specific parametrization of the two functions is described in the Supplemental Material~\cite{Supplemental}.

\begin{figure}
\begin{center}
\includegraphics[width=8.0cm]{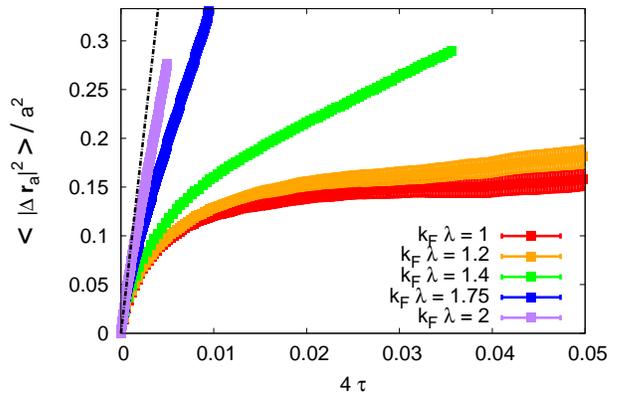}
\caption{(color online). Mean square displacement, in units of the lattice spacing $a$, of the impurity coupled to the WC phase at $k_Fr_0 = 35$. The imaginary time $\tau$ is in units of $mr_0^2/\hbar$. The dashed line corresponds to the free diffusion $m/m^\ast=1$.}
\label{fig3}
\end{center}
\end{figure}

Results will be presented first for the WC phase and then for the FL phase:

\underline{\it{Impurity coupled to a Wigner crystal.}}
\textemdash$\;$ The main results of the FNDMC simulations are reported in Figs.~\ref{fig1}-\ref{fig2}. In Fig.~\ref{fig1} we show the binding energy $|\mu|$ as a function of the interlayer distance $k_F\lambda$ for three values of the dipole-dipole interaction 
strength. The value of $|\mu|$ varies by orders of magnitude and furthermore, when scaled in units of $\frac{\epsilon_F}{2}k_Fr_0$, the results practically overlap showing a very small dependence on the value of $k_Fr_0$. At short distances we find agreement with the energy of the two-body bound state of the potential (\ref{V}), present for any value of $k_F\lambda$~\cite{Simon76, Klawunn10, Armstong10}. In Fig.~\ref{fig2} we report the results of the inverse effective mass for the same three values of $k_Fr_0$ shown in Fig.~\ref{fig1}. By reducing the distance $\lambda$ from $3/k_F$ to $1/k_F$ the effective mass changes from a free regime, where $m/m^\ast\sim1$, to values $m/m^\ast\ll1$. For example, at $k_F\lambda=1$, we find $m/m^\ast=0.006(2)$ for $k_Fr_0=35$. The increase of the mean square displacement of the impurity, $|\Delta {\bf r}_a(\tau)|^2$, as a function of imaginary time is shown in Fig.~\ref{fig3} in the case $k_Fr_0=35$. The results for $m/m^\ast$ are obtained, following Eq.~(\ref{Diff}), by fitting a line to the long-time behavior of these curves. The dramatic increase of the effective mass at small values of $k_F\lambda$ is clearly shown by the tendency of the long-time tail to approach an horizontal line.

Direct contact with the acoustic polaron model can be made when the value of $k_Fr_0$ is deep enough in the WC phase that the harmonic approximation for the single-layer Hamiltonian is valid~\cite{Matveeva12,Mora07}. By expanding the particle-impurity interaction term in a sum over the excitations of the lattice, similarly to the derivation of the electron-phonon interaction in crystals~\cite{MahanBook}, one can write the Hamiltonian of the bilayer system as: 
\begin{eqnarray}
&& H = U_0+\sum_{{\bf q},s}\hbar\omega_{{\bf q},s}\left( a_{{\bf q},s}^\dagger a_{{\bf q},s}+\frac{1}{2}\right) -\frac{\hbar^2}{2m}\nabla_a^2 + U({\bf r}_a)
\nonumber
\\
&&+ \frac{i}{\Omega} \sum_{{\bf q}, s} V_{\bf q} \, e^{i {\bf q} \cdot {\bf r}_a} {\bf q} \cdot {\bf e}^\ast_{{\bf q},s}\sqrt{\frac{\hbar N}{2 m \omega_{{\bf q},s}}} (a_{{\bf q}, s} + a^\dagger_{-{\bf q},s}) \;. 
\label{Hamilton1}
\end{eqnarray}
Here $U_0=1.597\frac{k_Fr_0}{8}N\epsilon_F$ is the energy of the lattice in the classical limit~\cite{Mora07}. $U({\bf r}_a)=\sum_{m=1}^NV(|{\bf R}_m-{\bf r}_a|)$ is the static periodic potential when the atoms occupy the lattice sites ${\bf R}_m$, whose spatial average over the primitive cell is vanishing. Sums run over the wavevectors ${\bf q}$ of the first Brillouin zone and over the two branches $s$, corresponding to phonons with energy $\hbar\omega_{{\bf q},s}$ whose creation and annihilation operators are denoted by $a_{{\bf q},s}^\dagger$ and $a_{{\bf q},s}$ respectively. The interlayer potential (\ref{V}) enters the above equation with its Fourier transform $V_{\bf q} = - 2 \pi d^2 q e^{-q\lambda}$ and ${\bf e}_{{\bf q},s}$ denotes the polarization unit vector obeying to ${\bf e}_{{\bf q},s}^\ast\cdot{\bf e}_{{\bf q},s^\prime}=\delta_{s,s^\prime}$. We notice that higher-order phonon terms as well as {\it umklapp} processes are neglected in the Hamiltonian (\ref{Hamilton1}). 

Perturbation theory can be applied to the Hamiltonian (\ref{Hamilton1}) in the limit of a weak interlayer coupling potential $V_{\bf q}$. The increase in energy with respect to the unperturbed ground-state $E_{\bf k}=U_0+\frac{1}{2}\sum_{{\bf q},s}\hbar\omega_{{\bf q},s}+\frac{\hbar^2k^2}{2m}$ for an impurity moving with a small momentum $\hbar{\bf k}$ is given by $\delta E_{\bf k}=\mu+\frac{\hbar^2k^2}{2}(\frac{1}{m^\ast}-\frac{1}{m})$ and allows one to determine both the binding energy and the effective mass. The contribution from the static potential $U({\bf r}_a)$ is exponentially suppressed at large $k_F\lambda$ and can be neglected. The coupling to phonons gives instead the result
\begin{equation}
\mu=- \epsilon_F \frac{0.455}{(k_F\lambda)^4} k_Fr_0 \;,
\label{perturbation1}
\end{equation}
for the binding energy and
\begin{equation}
\frac{m^\ast}{m}  = 1+\frac{0.553}{(k_F\lambda)^4} \;,
\label{perturbation2}
\end{equation}
for the effective mass. The derivation of results (\ref{perturbation1})-(\ref{perturbation2}) makes use of the excitation energy $\hbar\omega_{{\bf q},\ell}=c_\ell q$ of longitudinal long-wavelength phonons, where $c_\ell=0.642\sqrt{k_Fr_0}\frac{\hbar k_F}{m}$ is the corresponding speed of sound obtained numerically using the approach of Ref.~\cite{Mora07}.

The above predictions of perturbation theory are compared with QMC results in Fig.~\ref{fig1} and in Fig.~\ref{fig2}, respectively for $\mu$ and $m/m^\ast$. In the case of the binding energy, a good agreement is found when $k_F\lambda\gtrsim4$ (see Fig.~\ref{fig1}). For large values of $k_F\lambda$ the increase of the effective mass is such a tiny effect that the limited accuracy of the QMC results does not allow for a useful comparison with Eq.~(\ref{perturbation2}). Instead, as Fig.~\ref{fig2} clearly shows, $m/m^\ast$ is found to be appreciably smaller than unity for values of the coupling where Eq.~(\ref{perturbation2}) is still very close to the unperturbed value. 

In order to understand better the role of phonons, we performed calculations of $m/m^\ast$ using the single-particle Hamiltonian $-\frac{\hbar^2}{2m}\nabla^2_a+U({\bf r}_a)$ and we thereby determined the inverse band mass in the static potential $U({\bf r})$. In these simulations the particles in the bottom layer are considered to be fixed in the positions ${\bf R}_m$ of the Bravais lattice and phonon excitations are thus completely frozen. The results are reported in Fig.~\ref{fig2} for the value $k_Fr_0=35$ of the in-plane coupling strength. We notice that the suppression of $m/m^\ast$ with decreasing distance $k_F\lambda$ is much more pronounced in the static case than when quantum fluctuations are included. For $k_F\lambda\lesssim3$, these fluctuations produce a significant decrease of the effective mass, thus enhancing the ratio $m/m^\ast$, in contrast to what is typically expected from the coupling to phonon excitations. A possible physical explanation of this effect is phonon-assisted hopping of the impurity between lattice sites, which results in a reduction of the impurity effective mass. This mechanism competes with the increase of the effective mass arising from the phonon drag and becomes dominant for small enough values of $k_F\lambda$. 

\begin{figure}
\centering
\subfigure[]{
\includegraphics[height=4.4cm]{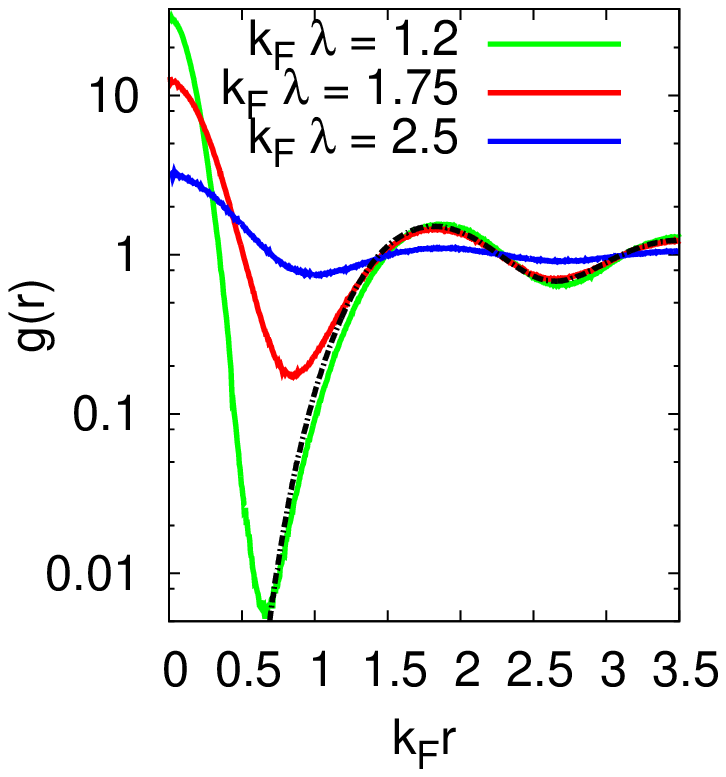}
\label{fig:subfig1_1}
}
\subfigure[]{
\includegraphics[height=4.4cm]{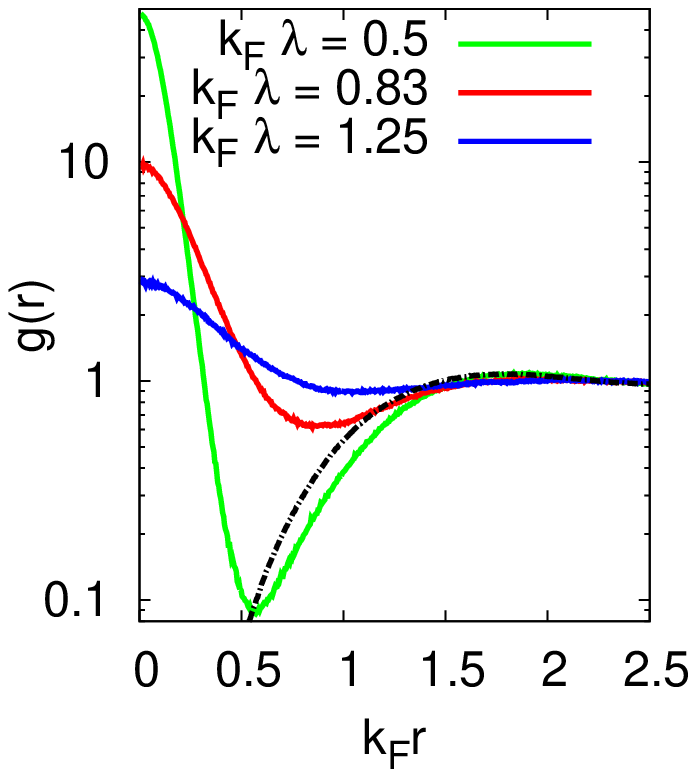}
\label{fig:subfig1_2}
}
\caption[Optional caption for list of figures]{(color online). Particle-impurity distribution function in terms of the in-plane distance. WC phase at  $k_Fr_0 = 35$ [panel \subref{fig:subfig1_1}] and FL phase at $k_Fr_0=2.5$ [panel \subref{fig:subfig1_2}]. The particle-particle correlation function of the clean system is also shown for the two cases (dashed line).}
\label{fig4}
\end{figure}

Finally in Fig.~\ref{fig:subfig1_1} we show the results for the particle-impurity correlation function, related to the probability of finding the impurity and a particle at a distance $r$ apart.
When $k_F\lambda$ is large, the impurity is highly mobile and at distances $k_Fr>1$ it experiences a uniform medium. As the interlayer separation is reduced, the large-distance tail of the distribution coincides with the particle-particle correlation
 function showing the structure of the WC phase~\cite{Matveeva12}. At the same time the peak at short distance becomes higher and a hole deepens at $k_Fr\lesssim1$ due to the in-plane dipolar repulsion. The structure of the crystal around the impurity is 
only slightly changed, with peaks that are a few percent higher, compared to the clean system.

\underline{\it{Impurity coupled to a Fermi liquid.}}

The FNDMC results for $|\mu|$ and $m/m^\ast$ when $k_Fr_0<(k_Fr_0)_c$, corresponding to the dipolar fermions in the bottom layer in the FL phase, are reported in Figs.~\ref{fig1}-\ref{fig2}. For the smallest value of the interaction strength, $k_Fr_0=0.5$, we find agreement at large interlayer separation with the binding energy obtained using a perturbation treatment based on a free Fermi gas~\cite{Klawunn13}. At short distance, instead, one always recovers the energy of the two-body bound state. The inverse effective mass is shown to decrease with increasing interlayer coupling strength (see Fig.~\ref{fig2}) but, in contrast to the WC phase, it approaches a finite asymptotic value for small $k_F\lambda$. If $k_Fr_0\ll1$ we expect in this regime $m^\ast\simeq2m$, the mass of a dimer. Interaction effects in the bottom layer make the value of the effective mass as large as $m^\ast\simeq4m$ at $k_Fr_0=20$. The particle-impurity correlation function is reported in Fig.~\ref{fig:subfig1_2}. It behaves qualitatively similar to the WC case with a peak caused by the particle-impurity attraction and a hole due to the in-plane repulsion. The main difference is the more pronounced deformation of the medium around the impurity in the FL phase.

An important issue concerns the experimental realizability of the dipolar bilayer system in the proper regime of parameters where strongly-coupled polarons can be investigated. 2D Fermi gases have been realized~\cite{Froehlich11,Orel11} with inverse Fermi wavevectors in the range $1/k_F\sim100-500$ nm. Optical lattices with high barriers and lattice spacing on the order of $500$ nm are also available. By using polar molecules produced with mixtures of $^{23}$Na-$^{40}$K~\cite{MIT12} or of $^{133}$Cs-$^6$Li~\cite{Heidelberg13} the dipolar length can be as large as $r_0\sim6.8-62$ $\mu$m and large values of $k_Fr_0$ can be achieved. If such molecules can be produced in their ground state and brought to quantum degeneracy, ultracold dipolar systems can become a useful tool to investigate the rich physics of polarons in the strong-coupling regime.

Useful discussions with G. Astrakharchik and A. Pikovski are gratefully aknowledged. This work has been supported by ERC through the QGBE grant and by Provincia Autonoma di Trento.  We thank the AuroraScience project (funded by PAT and INFN) for allocating part of the computing resources for this work and for technical support.

\end{document}


\title{The impurity problem in a bilayer system of dipoles: Supplemental Material}
\author{Natalia Matveeva and Stefano Giorgini}
\affiliation{Dipartimento di Fisica, Universit\`a di Trento and CNR-INO BEC Center, I-38050 Povo, Trento, Italy}

\maketitle

\section{Treatment of the dipole-dipole interaction energy}

Since the dipole-dipole force is long range, the potential energy contributions arising from in-plane ($V_{dd}$) and interlayer ($V_{ad}$) interactions require a careful treatment. The in-plane contribution is given by 
\begin{equation}
V_{dd}=\sum_{i<j}\frac{d^2}{|{\bf r}_i-{\bf r}_j|^3} + \frac{1}{2}\sum_{i,j}\sum_{{\bf R}\neq0} \frac{d^2}{|{\bf r}_i-{\bf r}_j-{\bf R}|^3} \;,
\label{Vdd}
\end{equation}
where $i$ and $j$ label particles of the bottom layer in the simulation cell and the vectors ${\bf r}_{i(j)}+{\bf R}$ correspond to the positions of all images of particle $i(j)$ in the array of replicas of the simulation cell. Similarly, the contribution from interlayer dipolar interactions is given by
\begin{equation}
V_{ad}=\sum_{i}\sum_{{\bf R}}\frac{d^2 (|{\bf r}_a-{\bf r}_i-{\bf R}|^2 -2\lambda^2)}{(|{\bf r}_a-{\bf r}_i-{\bf R}|^2 +\lambda^2)^{5/2}}\;,
\label{Vad}  
\end{equation}
where ${\bf r}_a$ denotes the projection onto the bottom layer of the impurity position. We calculate the mean interaction energy using a procedure that takes advantage of the fast $1/r^3$ decay of the dipole-dipole potential:
\begin{equation}
\langle V\rangle=(V_{dd})_{R_{c_1}}+E_{\text{tail}_1} + (V_{ad})_{R_{c_2}}+E_{\text{tail}_2} \;.
\label{Vaver}
\end{equation}
Here $(V_{dd})_{R_{c_1}}$ and $(V_{ad})_{R_{c_2}}$ denote the sums (\ref{Vdd}) and (\ref{Vad})  with the constraints $|{\bf r}_i-{\bf r}_j-{\bf R}|\le R_{c_1}$ and $|{\bf r}_a-{\bf r}_j-{\bf R}|\le R_{c_2}$, respectively. The corresponding tail contributions $E_{\text{tail}_1}=\pi nd^2/R_{c_1}$ and $E_{\text{tail}_2}=2\pi nR_{c_2}^2/(\lambda^2+R_{c_2}^2)^{3/2}$ are obtained by assuming a uniform distribution of particles for distances larger than the cut-off range.

\section{Particle-impurity and particle-particle Jastrow factors}
The trial wavefunction used in the QMC simulations has the general Jastrow-Slater form
\begin{equation}
\psi_T({\bf r}_1,...,{\bf r}_N,{\bf r}_a)=\prod_{i=1}^N h(r_{ai}) \prod_{i<j}f(r_{ij})\det[\varphi({\bf r}_i)]\;,
\label{psiT}
\end{equation}
where the determinant, constructed with $N$ single-particle states $\varphi({\bf r})$, fulfills the antisymmetry condition under exchange of the particles in the bottom layer and determines the position of the nodes of the many-body wavefunction. The first two factors in Eq.~(\ref{psiT}) consist of products, over all particles and all different pairs of particles in the bottom layer, of two-body correlation terms: $h(r)$ refers to the particle-impurity correlation and depends on their separation distance $r_{ai}=|{\bf r}_a-{\bf r}_i|$, $f(r)$ refers instead to the particle-particle correlation, being $r_{ij}=|{\bf r}_i-{\bf r}_j|$ the corresponding pair distance. Both $h(r)$ and $f(r)$ are non-negative, spherically symmetric functions parametrized in the following way: 
\begin{eqnarray}
h(r)= \begin{cases}
          e^{-\gamma r^2}   &\text{if $r<R_1$\;;} \\
          \frac{C}{2}\left[ e^{-\beta(r-R_2)}+e^{\beta(r-R_2)}\right]  &\text{if $R_1<r<R_2$\;;}\\
          C &\text{if $r>R_2$\;.}
          \end{cases}
\label{hr}
\end{eqnarray}
The coefficients $C$, $\gamma$ and $\beta$ ensure continuity of $h(r)$ at $r=R_1$ and of the derivative $h^\prime(r)$ at $r=R_1$ and $r=R_2$. In this case correlations are modeled by a Gaussian at short distance and by a flat plateau at large distance. The matching points $R_1$ and $R_2$, with $R_1<R_2$, are variational parameters that should be optimized in order to reduce the statistical variance of the calculation. For the particle-particle Jastrow term we use instead the following form:
\begin{eqnarray}
f(r)=\begin{cases}
        K_0(\frac{2}{\sqrt{r}})  &\text{if $r<R_{\text{match}}$\;;} \\
        C_1\exp(-\frac{C_2}{r(L-r)}) &\text{if $r>R_{\text{match}}$\;.}
        \end{cases}
\label{fr}
\end{eqnarray}
Here $K_0(r)$ is the modified Bessel function and $C_1$ and $C_2$ are again constants enforcing the continuity of $f(r)$ and $f^\prime(r)$ at the matching point $r=R_{\text{match}}$. The value of $R_{\text{match}}$ is a variational parameter used to reduce the statistical variance. The choice of the short-range part of $f(r)$ arises from the cusp condition with the $1/r^3$ potential, while the exponential for $r>R_{\text{match}}$ is a suitable functional form which satisfies the boundary condition $f^\prime(r=L/2)=0$, being $L$ the side of the square simulation box. In the WC phase, when a rectangular simulation box of volume $\Omega=L_xL_y$ is used in order to accommodate an integer number of primitive cells of the Bravais lattice, the length $L$ in Eq.~(\ref{fr}) is replaced by the smallest of the two sides $L_x\le L_y$. An identical choice for the particle-particle Jastrow correlation was used in the study of Ref.~\cite{Matveeva07}.